\documentclass[a4paper, 11pt]{article}

\usepackage{bm}
\usepackage{setspace}
\usepackage{graphicx}
\usepackage{xcolor}
\usepackage{slashed}
\usepackage{ amssymb }
\usepackage{textgreek}
\usepackage{mathtools}
\usepackage{latexsym}
\usepackage{textgreek}
\usepackage{verbatim}
\usepackage{xparse,mathtools}
 \usepackage[utf8]{inputenc}
 \usepackage{amsmath}
\usepackage{amsfonts}
\usepackage{verbatim}
\usepackage{graphicx}
\usepackage{subfigure}
\usepackage{amssymb}
\usepackage[T1]{fontenc}
\usepackage{slashed}
\usepackage{color}

\usepackage[numbers,sort&compress]{natbib}
\usepackage[linktocpage=true,hidelinks]{hyperref}

\def\beq{\begin{eqnarray}}
\def\eeq{\end{eqnarray}}

\def\({\left(}
\def\){\right)}

\newcommand{\be}{\begin{equation}}
\newcommand{\ee}{\end{equation}}
\newcommand{\la}{\langle}
\newcommand{\ra}{\rangle}
\def\ea{\end{eqnarray}}
\def\ba{\begin{eqnarray}}

\def\beq{\begin{eqnarray}}
\def\eeq{\end{eqnarray}}

\def\({\left(}
\def\){\right)}
\def\mn{_{\mu \nu}}

\def\p{\partial}
\def\la{\langle}
\def\ra{\rangle}

\def\lsim{\mathrel{\rlap{\lower3pt\hbox{\hskip0pt$\sim$}}
     \raise1pt\hbox{$<$}}}         
\def\gsim{\mathrel{\rlap{\lower4pt\hbox{\hskip1pt$\sim$}}
     \raise1pt\hbox{$>$}}}         

\def\lsim{\mathrel{\rlap{\lower3pt\hbox{\hskip0pt$\sim$}}
     \raise1pt\hbox{$<$}}}         
\def\gsim{\mathrel{\rlap{\lower4pt\hbox{\hskip1pt$\sim$}}
     \raise1pt\hbox{$>$}}}         

\begin{document}

\def\thefootnote{\fnsymbol{footnote}}

\begin{center}
{\Large \bf{Perturbative Construction of Coherent States}}

 \vspace{1truecm}
\thispagestyle{empty} \centerline{\large  {Lasha  Berezhiani~$^{1,2}$\footnote{lashaber@mpp.mpg.de}, Giordano Cintia~$^{1,2}$\footnote{cintia@mpp.mpg.de} and Michael Zantedeschi~$^3$}\footnote{zantedeschim@sjtu.edu.cn}
}

 \textit{$^1$Max-Planck-Institut f\"ur Physik, F\"ohringer Ring 6, 80805 M\"unchen, Germany\\
 \vskip 5pt
$^2$ Arnold Sommerfeld Center, Ludwig-Maximilians-Universit\"at, \\Theresienstra{\ss}e 37, 80333 M\"unchen, Germany\\
 \vskip 5pt
$^3$ Tsung-Dao Lee Institute and School of Physics and Astronomy,\\
    Shanghai Jiao Tong University, Shengrong Road 520, 201210
    Shanghai, China
 }

\end{center}  

 \def\thefootnote{\arabic{footnote}}
\setcounter{footnote}{0}
\begin{abstract}
The perturbative consistency of coherent states within interacting quantum field theory requires them to be altered beyond the simple non-squeezed form. Building on this point, we perform explicit construction of consistent squeezed coherent states, required by the finiteness of physical quantities at the one-loop order. Extending this analysis to two-loops, we demonstrate that a non-Gaussian alteration of squeezed coherent states is necessary. The modifications of the coherent state we propose are perturbative in $\hbar$ and may be an indication that coherence must be viewed through a nonlinearly redefined, background-dependent, degree of freedom.

\end{abstract}

\newpage
\setcounter{page}{1}

\renewcommand{\thefootnote}{\arabic{footnote}}
\setcounter{footnote}{0}

\linespread{1.1}
\parskip 4pt

\section{Introduction}

Coherent states are recognized as the quantum counterparts of classical configurations~\cite{Glauber:1963tx,Sudarshan:1963ts,Kibble:1965zza} (see also~\cite{Zhang:1990fy,Zhang:1999is}), which was precisely Schr\"odinger's motivation for introducing them.
However, their utilization has been limited to systems for which non-linearities do not play a significant role.  The scope and implications of the coherent state representation for various nonlinear classical systems within the quantum field theoretic framework started emerging a little over a decade ago, in the series of papers~\cite{Dvali:2011aa,Dvali:2012en,Dvali:2012wq,Dvali:2013eja,Dvali:2013vxa}. There, this class of states is proposed as the prime candidate for the underlying quantum description of classical backgrounds comprised of interacting constituents.

For most of the classical systems, the implementation of this framework is unnecessary, since the timescales over which the configuration can be thought of as an isolated one (even classically) tend to be significantly shorter than the time it takes for quantum effects to become important. The timescale for the significant quantum departure from the classicality was coined as \textit{quantum break time} in~\cite{Dvali:2013vxa}. This phenomenon was explored further in~\cite{Dvali:2013eja,Dvali:2013lva,Berezhiani:2016grw,Dvali:2017eba,Vachaspati:2017jtw,Dvali:2017ruz,Kovtun:2020ndc,Kovtun:2020udn,Berezhiani:2020pbv,Berezhiani:2021gph,Dvali:2022vzz,Michel:2023ydf}. The argument is based on the idea that classical configurations must be viewed fundamentally as quantum 
states with high occupation numbers of aligned modes, such as condensates or coherent states.   
As a result, quantum interactions between these modes lead to the departure from coherence and thereby from classicality. 
It was shown that the quantum break time is generically expected to increase for weaker quantum couplings. 
In particular, it scales as a negative power of the coupling for classically-stable systems\footnote{ The scaling in terms of negative powers of the quantum coupling $\lambda$ can be converted as a positive-power scaling in the occupation number $n$, for a fixed collective coupling $\alpha=\lambda n$.  The latter is the parameter controlling the magnitude of the classical non-linearities of the theory.},
while it becomes logarithmic if the system exhibits a classical Lyapunov instability, with the coefficient set by the Lyapunov exponent~\cite{Dvali:2013vxa,Dvali:2015wca,Kovtun:2020udn}.

Furthermore, it has been demonstrated that the fundamental quantum description in terms of coherent states can be of utmost importance for classically (quasi-)eternal systems, such as black hole\footnote{For systems such as black holes, the phenomenology of the quantum breaking can be significantly altered due to high micro-state degeneracy, because of the so-called memory burden effect~\cite{Dvali:2018xpy,Dvali:2020wft}.}, de Sitter and inflationary spacetimes~\cite{Dvali:2011aa,Dvali:2012en,Dvali:2012wq,Dvali:2013vxa,Dvali:2013eja,Dvali:2013lva,Berezhiani:2016grw,Dvali:2017eba}. 
 The implications of these concepts for cosmic axions~\cite{Dvali:2017ruz}, ultra-light scalar dark matter models~\cite{Eberhardt:2023axk} and the onset of inflation~\cite{Berezhiani:2015ola,Berezhiani:2022gnv} have also been considered. See \cite{Hertzberg:2016tal}, for a related interesting discussion of the classicality in interacting bosonic systems.

Commonly, quantum backreaction on classical dynamics is analyzed within the so-called background field method. With this approach, non-equilibrium quantum processes are studied by analyzing the coupled evolution of the background field, represented by the one-point expectation value of the field operator, and higher order correlation functions, see e.g.~\cite{Berges:2004yj} and references therein. Already within this approach, some of the quantum effects can be captured successfully. A good example is provided by a quartic scalar field theory, within which the homogeneous anharmonic oscillations tend to depart from a classical trajectory increasingly quickly with increasing amplitude of oscillations~\cite{Baacke:1996se}. For a relatively recent interesting work on a related subject see also~\cite{Ai:2021gtg}. The background field method has been also employed recently to analyze the stability of the homogeneous non-Abelian electric field background ~\cite{Pereira:2022lbl,Vachaspati:2022gco}.

However, the piecewise treatment of the state merely in terms of a few correlation functions may obscure the ramifications of some effects for the state of the system in its entirety. Therefore, since coherent states are the prime candidates for the full quantum description of classical backgrounds as discussed above, their self-consistent construction within interacting quantum field theory is of fundamental importance.

To illustrate the point, let us begin with the simplest family of coherent states constructed from a real scalar quantum field $\hat{\phi}(x,t)$. These states, known as \textit{non-squeezed coherent states}, are typically characterized by a set of c-number functions ${\phi_{\rm cl}(x),\pi_{\rm cl}(x)}$ that determine the initial conditions for the system. Their general form can be expressed as follows
\beq
|C\ra=\exp\left\{-{\rm i}\int {\rm d}^3 x \left( \phi_{\rm cl}(x)\hat{\pi}(x)-\pi_{\rm cl}(x)\hat{\phi}(x) \right)\right\}|\Omega\ra,
\label{intro:coherent}
\eeq
where $\hat{\pi}(x,t)$ is the conjugate momentum of the field operator, satisfying standard canonical commutation relations,
and $|\Omega\ra$ is the vacuum of the interacting theory\footnote{In this paper we work in units of $\hbar =c=1$.}. Initial conditions for the expectation values of the field and its conjugate momentum are uniquely determined by the c-number functions as 
\beq
\label{initphi}
&&\la C | \hat{\phi} |C\ra (t=0)=\phi_{\rm cl}(x)\,,\\
&&\la C | \hat{\pi} |C\ra (t=0)=\pi_{\rm cl}(x)\,.
\label{initpi}
\eeq
If $\phi_{\rm cl}(x)$ and $\pi_{\rm cl}(x)$ remain finite as $\hbar$ approaches zero, the evolution of the state \eqref{intro:coherent} in this limit will be completely captured by the replacement of these c-number functions with time-dependent classical configurations satisfying the same initial conditions\footnote{If we were to reintroduce $\hbar$ explicitly, we would have to add a factor of $\hbar^{-1}$ in the exponent of \eqref{intro:coherent} for this property to be fulfilled.
}.

Constructing the state in terms of the operators of the fully interacting theory rather than their asymptotic counterparts offers certain advantages within the quantum field theory framework. In particular, this approach enables computations to be carried out without immediate reliance on perturbation theory. This is made possible by the application of canonical commutation relations, which allow for efficient manipulations of the operators and the vacuum involved in the construction of the state. Consequently, the transition to the interaction picture can be postponed to a later stage, when the exact computation reaches its limits and the necessity for a loop expansion becomes apparent.

It was demonstrated in~\cite{Berezhiani:2020pbv,Berezhiani:2021gph} that non-squeezed coherent states exhibit unrenormalizable perturbative divergencies within interacting quantum field theories. This phenomenon bears resemblance to certain initial-time singularities encountered in the semi-classical analysis of backreaction from fluctuations to the background~\cite{Baacke:1997zz,Baacke:1999ia} in the context of scalar field theories with quartic interactions. 
To understand the source of these divergences, it is important to understand how non-squeezed coherent states determine the initial conditions for the two-point and higher-order correlation functions. At the initial time, the two-point correlation functions of canonical degrees of freedom on the state $|C\rangle$ read
\beq
\label{init2phi}
&&\la C | \hat{\phi}(x,0)\hat{\phi}(y,0) |C\ra=\phi_{\rm cl}(x)\phi_{\rm cl}(y)+\la \Omega | \hat{\phi}(x,0)\hat{\phi}(y,0) |\Omega\ra\,,\\
&&\la C | \hat{\pi}(x,0)\hat{\pi}(y,0) |C\ra=\pi_{\rm cl}(x)\pi_{\rm cl}(y)+\la \Omega | \hat{\pi}(x,0)\hat{\pi}(y,0) |\Omega\ra\,.
\label{init2pi}
\eeq
Notice that we have not specified a theory yet, nevertheless the expressions for initial conditions \eqref{initphi}-\eqref{init2pi} are exact and merely rely on the fact that one-point expectation values vanish on the vacuum.

The reason \eqref{init2phi} and \eqref{init2pi} are interesting is because the fluctuation part is given by the vacuum two-point function and is independent of the background, i.e. of $\phi_{\rm cl}(x)$ and $\pi_{\rm cl}(x)$. We anticipate that it is exactly this property that fails in giving a good perturbative behaviour of the state. The main issue is that initial conditions for correlators, when not specified at an asymptotic time, should satisfy a minimal amount of dressing and should provide certain divergences when evaluated at coincidence. Only in this way, the dynamics is renormalizable in the conventional sense. Unfortunately, this minimal tuning cannot be satisfied by the simplest class of coherent states. 

At this point, it is crucial to highlight the nature of these divergences.
To demonstrate the failure of this state in providing a perturbative dynamics, we can focus on the theory of a massive scalar field with quartic interactions, governed by the following Hamiltonian
\beq
\hat{H}=\int {\rm d}^3x\left[ \frac{1}{2}\hat{\pi}^2+ \frac{1}{2}\left(\vec{\nabla}\hat{\phi} \right)^2+\frac{1}{2}m^2\hat{\phi^2}+\frac{\lambda}{4!}\hat{\phi}^4\right]\,.
\label{quarticH}
\eeq
Here $m$ and $\lambda$ are the bare mass and coupling constant, respectively. Note, we have not introduced the field renormalization explicitly as it equals to unity at one-loop order for the quartic theory in question, thus we keep it implicit for the time being and will manifest it when we move to higher-order loop corrections.

The application of Baker--Campbell--Hausdorff formula, and taking into account that the vacuum expectation values of an odd number of fields vanish for the quartic theory in question (see~\cite{Berezhiani:2020pbv}), straightforwardly leads to the following exact statements
\beq
\label{acceq}
\partial_t^2\la C|\hat{\phi}|C\ra(t=0)&=&\Delta \phi_{\rm cl}-\phi_{\rm cl} \left(m^2+\frac{\lambda}{2}\la\Omega|\hat{\phi}^2|\Omega\ra\right)-\frac{\lambda }{3!}\phi_{\rm cl}^3\,,\\
\la C|\hat{H}|C \ra &=&\int {\rm d}^3x \left[ \frac{1}{2}\pi_{\rm cl}^2+\frac{1}{2}\left(\vec{\nabla}\phi_{\rm cl} \right)^2+\frac{1}{2}\left( m^2+\frac{\lambda }{2}\la \Omega|\hat{\phi}^2|\Omega\ra \right)\phi_{\rm cl}^2 +\frac{\lambda }{4!}\phi_{\rm cl}^4 \right]\,.
\label{intro:energy}
\eeq
These are the initial acceleration of the one-point function of the field and the energy of the system. It is important to emphasise that we have adjusted the bare vacuum energy in a way that the expectation value \eqref{intro:energy} vanishes if the c-number functions are sent to zero. 

The bare mass term $m^2$ comes with the entourage of the divergent contribution $\langle \Omega |\hat{\phi}^2|\Omega\rangle$, which corresponds to the so-called \textit{bubble diagram} and represents the same divergent loop correction that arises from the standard analysis of quantum corrections to the vacuum propagator at one-loop. However, the analogous $S$-matrix contribution to the coupling constant is absent in equations \eqref{acceq} and \eqref{intro:energy}. In other words, the energy and initial field acceleration of a non-squeezed coherent state are sourced by the bare coupling constant, and not by its physical counterpart.

This discrepancy between the divergences provided by a state at the initial time and the standard $S$-matrix divergences is commonly referred to as the \textit{Initial Time Singularity}~\cite{Baacke:1997zz,Baacke:1999ia}. The name arises from the observation that, when we examine the evolution of the expectation value of the field operator, we discover that the missing $S$-matrix singularities are perturbatively generated at $t>0$. However, these singularities are not initially present and, through the process of renormalization, they are effectively shifted to the moment of initialization of the coherent state. Because of this issue, non-squeezed coherent states as given by \eqref{intro:coherent} can only be considered as physical states of the theory if the required divergences that are not accounted for within perturbation theory would somehow resum into finite results. For instance, the finiteness of the bare coupling constant appears to be essential for the validity of these states.

However, if we would like to maintain physical quantities to be perturbatively finite, we need to abandon the simplicity of these states and instead allow for their nontrivial squeezing and more. We will introduce the corresponding procedure in detail starting from the next sections. As an appetizer, we already mentioned that the root of all trouble comes from the quadratic correlation function \eqref{init2phi} and the fact that its fluctuation part is independent of $\phi_{\rm cl}$. Therefore, it appears to us that the only possibility forward is to build the coherent state around an excited state instead of the vacuum, i.e. 
\beq
|C\ra=\exp\left\{-{{\rm i}}\int {\rm d}^3 x \left( \phi_{\rm cl}(x)\hat{\pi}(x)-\pi_{\rm cl}(x)\hat{\phi}(x) \right)\right\}|S\ra\,,
\label{intro:squeezed}
\eeq
where $|S\ra$ is the squeezed vacuum with the one-point function of the field operator vanishing in this state. In the next section, we will begin the analysis with $\phi_{\rm cl}=\phi_0=\text{const
}$ and $\pi_{\rm cl}=0$ for simplicity. Also, the homogeneous backgrounds are the ones usually studied within the semi-classical framework and we can use the intuition gained from those studies to fast-track the computation.

The paper is organized as follows. In Sec.\,\ref{sec:1-loop}, we derive the squeezed coherent state which is capable of addressing the initial time singularity at $\hbar$-order. In Sec.\,\ref{sec:Tstate}, we discuss the emergence of initial time singularities at $\hbar^2$-order. We show that squeezed coherent states fall short of addressing them at this order and that a non-Gaussian generalization is required.  In  Sec.\,\ref{sec:gauge}, we discuss initial time singularities in scalar electrodynamics. We summarize the results in Sec.\,\ref{sec:summary}.

\section{Squeezed Coherent states and one-loop dynamics}
\label{sec:1-loop}

As we have already mentioned, the trouble stems from \eqref{init2phi} and the fact that its fluctuation part is independent of $\phi_{\rm cl}$. This point was already noticed in the semiclassical analysis and the issue was often addressed by proposing a specific way to fix initial conditions for fluctuations~\cite{Baacke:1997zz}.

For simplicity, let us focus our discussion here on a homogeneous scalar field configuration $\phi_0$ and the corresponding quantum state within the theory governed by \eqref{quarticH}. From the analysis of the effective potential for classically stationary backgrounds \textit{á la} Coleman-Weinberg, say in metastable vacuum case or spontaneous symmetry breaking scenario, we are accustomed to the background dependent dynamical equations for quantum fluctuations $\delta\hat{\phi}=\hat{\phi}-\phi_0$ (see e.g.~\cite{mukhanov})
\beq
\left(-\Box+m^2+V''(\phi_0)\right)\delta\hat{\phi}=0\,,
\label{constphieq}
\eeq
indicating that the presence of the background $\phi_0$ introduces a background-dependent contribution to the mass of excitations
\beq
m^2_{\rm exc}=m^2+V''(\phi_0)\,.
\eeq
In this case, a natural choice for the vacuum of $\delta\hat{\phi}$ would be with respect to the mode functions of a massive particle with mass $m_{\rm exc}$, so that the initial conditions for correlation functions depend on $\phi_0$.
Within the theory at hand \eqref{quarticH}, on the other hand, an initial state with homogeneous $\phi_0$ is classically oscillating. Therefore, the equation for fluctuations \eqref{constphieq} changes to
\beq
\left(-\Box+m^2+\frac{\lambda}{2}\phi_{\rm cl}^2(t)\right)\delta\hat{\phi}=0\,,
\label{oscphieq}
\eeq
with $\phi_{\rm cl}(t)$ being the solution to the classical equation of motion with initial conditions $\phi_{\rm cl}(0)=\phi_0$ and $\dot{\phi}_{\rm cl}(t)=0$. Obviously, the fluctuation equation receives a time-dependent effective mass $m_{\rm eff}^2(t)\equiv m^2+\lambda\phi_{\rm cl}^2(t)/2$. Thus, there is no notion of a constant effective mass here.

Semiclassical analysis shows that if the initial condition of quantum fluctuations in the vacuum is defined with respect to the background independent mode functions with mass $m$, then the backreaction from these quantum fluctuations on the background dynamics is singular. This so-called initial-time singularity is absent if the initial conditions for the fluctuation mode functions are chosen to be of the massive particle with mass $m_{\rm exc}=m_{\rm eff}(0)$ instead of $m$. This amounts to a Bogoliubov transformation of the mode functions. All these were illustrated in great detail in~\cite{Baacke:1997zz,Baacke:1999ia}.

This raises the question of what this implies in terms of the state of the system. It is well-known that different vacua of the ladder operators related by the Bogoliubov transformation have nontrivial particle content with respect to each other and are connected by the so-called squeezing operator; see e.g.~\cite{Mukhanov:2007zz}. Therefore, an intelligent guess for making the coherent state description of classical backgrounds perturbatively finite is to consider \textit{squeezed} coherent states
\beq
|C_S\ra=\exp\left\{{-{\rm i}\int {\rm d}^3 x \, \phi_{0}\,\hat{\pi}(x)}\right\}|S\ra\,, \qquad \text{with} \quad |S\rangle= {\rm e}^{-{\rm i}\hat{S}}|\Omega\rangle
\label{hom:squeezed}
\eeq
where $|S\ra$ is the squeezed vacuum, with the one-point function of the field operator vanishing in this state.

Next, we demonstrate how the above-mentioned semiclassical remedy for the initial-time singularity translates in the coherent state picture, fixing the issues reported both in \eqref{acceq} and \eqref{intro:energy}. In the case of the homogeneous background, the initial conditions that have been adopted in the semiclassical analysis of~\cite{Baacke:1999ia}
 corresponds to the choice of the basis for our coherent state \eqref{hom:squeezed} to satisfy
\beq
\label{modinitphi}
&&\la S | \hat{\phi}(x,0)\hat{\phi}(y,0) |S\ra=\int \frac{{\rm d}^3 k}{(2\pi)^3} \frac{{\rm e}^{{\rm i}\vec{k}\cdot (\vec{x}-\vec{y})}}{2\sqrt{k^2+m^2+\frac{\lambda}{2}\phi_0^2}}\,,\\
&&\la S | \hat{\pi}(x,0)\hat{\pi}(y,0) |S\ra=\int \frac{{\rm d}^3k}{(2\pi)^3} ~\frac{1}{2}{\rm e}^{{\rm i}\vec{k}\cdot (\vec{x}-\vec{y})}\sqrt{k^2+m^2+\frac{\lambda}{2}\phi_0^2}\,,
\label{modinitpi}
\eeq
at least for the purposes of one-loop computations. Notice that instead of bare mass we could have had $m_{\rm ph}^2\equiv m^2+\frac{\lambda}{2}\la\Omega|\hat{\phi}^2|\Omega\ra$, incorporating the renormalization prescription absorbing the divergent vacuum bubble. However, for addressing the question of initial-time singularity, we will be predominantly utilizing the above expressions at the coincidence and therefore the difference between $m$ and $m_{\rm ph}$ would only manifest itself at higher loops. Of course, it will be relevant for the analysis of $(n\geq 2)$-point correlation functions. However, for the time being, we focus merely on the initial acceleration for the one-point function and the expectation value of the Hamiltonian.

The relevant question at this point concerns the explicit form of the squeezed vacuum $|S\ra$ that does the job. A class of relevant squeezing operations can be characterized as
\beq
|S\ra=\exp\left\{-{\rm i}\hat{S}\right\}|\Omega\ra\,, \qquad {\rm with} \qquad \hat{S}=\frac{1}{2}\int {\rm d}^3x \,{\rm d}^3y\, D(x-y) \left\{\hat{\phi}(x) \hat{\pi}(x)+\hat{\pi}(x) \hat{\phi}(y)\right\}\,,
\label{squeezeD}
\eeq
where $D(x-y)$ is a c-number function connected to the initial conditions we want to imprint in correlation functions. We demonstrate in appendix A that an explicit form of $D(x-y)$ reproducing initial conditions \eqref{modinitphi} and \eqref{modinitpi} is given by
\begin{equation}
     D(x-y)=-\frac{1}{4}\int\frac{{\rm d}^3 k}{(2\pi)^3}{\rm ln}\left(\frac{E_k}{\omega_k}\right){\rm e}^{{\rm i} k \cdot(x-y)}\,.
\label{eq:SqueezingPos}
\end{equation}
Here, we introduced the notation for dispersion relations $\omega_k\equiv\sqrt{k^2+m^2}$ and $E_k\equiv\sqrt{\omega_k^2+\frac{\lambda}{2}\phi_0^2}$. In other words, this is the choice that removes the initial time singularity at one-loop order. Notice that there is a plethora of squeezing operators for which the initial time-singularity is removed. However, the high momentum contribution to the integrand of \eqref{eq:SqueezingPos} must be universal.

Expansion of $\hat{S}$ in terms of creation and annihilation operators yields the following compact expression
\begin{equation}
    \hat{S}=\frac{{\rm i}}{4}\int \frac{{\rm d}^3k}{(2\pi)^3}\ln\left(\frac{E_k}{\omega_k}\right)\left(a_k a_{-k}-a_k^\dagger a_{-k}^\dagger\right)\,,
\label{eq:squeezing_parameter}
\end{equation}
which the reader might be more familiar with from quantum mechanics. The operator \eqref{eq:squeezing_parameter}, written in terms of ladder operators, was introduced in~\cite{Baacke:1997zz} to address the initial time singularity appearing at order $\hbar$ in the context of the semiclassical analysis.

As illustrated in Appendix A, the presence of the squeezed vacuum corresponds to implementing a Bogoliubov transformation. When evaluating correlation functions, this transformation allows us to compute any correlator over the state $|C_S\rangle$ by replacing the field content with a rotated field
\begin{flalign}
 & 
\quad\label{eq:phis}\hat{\phi}_s={\rm e}^{{\rm i}\hat{S}}\hat{\phi}{\rm e}^{-{\rm i}\hat{S}} =\int\frac{{\rm d}^3 k}{(2\pi)^3\sqrt{2E_k}}\left( \hat{a}_k  {\rm e}^{{\rm i}k \cdot x}+ \hat{a}^\dagger_k {\rm e}^{-{\rm i}k \cdot x} \right)\,,\\
     &\,\hat{\pi}_s={\rm e}^{{\rm i}\hat{S}}\hat{\pi}\,{\rm e}^{-{\rm i}\hat{S}}=(-{\rm i})\int\frac{{\rm d}^3 k}{(2\pi)^3}\sqrt{\frac{E_k}{2}}\left( \hat{a}_k  {\rm e}^{{\rm i}k \cdot x}- \hat{a}^\dagger_k {\rm e}^{-{\rm i} k \cdot x} \right)\,.
     \label{eq:pis}
\end{flalign}
The correlation functions of these new operators are then evaluated on the true vacuum $|\Omega\rangle$. It is important to note that the choice of this particular squeezing initiates mode functions in such a way that avoids mixing $\pm E_k$ frequency modes, otherwise unavoidable in the absence of squeezing. Also, the relations \eqref{eq:phis} and \eqref{eq:pis} imply that the effect of squeezing can be resolved non-perturbatively through a rotation of the operators\footnote{If $t>0$, the rotated operator $\hat{\phi}_s(x,t)$ reads
\begin{flalign}
{\rm e}^{{\rm i}\hat{S}(t_0)}\hat{\phi}(x,t){\rm e}^{-{\rm i}\hat{S}(t_0)} =\int\frac{{\rm d}^3 k}{(2\pi)^3\sqrt{2E_k}}\left\{\left(\cos \omega_k t-{\rm i}\frac{E_k}{\omega_k}\sin \omega_k t\right)\hat{a}_k  {\rm e}^{{\rm i}k \cdot x}+ \left(\cos \omega_k t+{\rm i}\frac{E_k}{\omega_k}\sin \omega_k t\right)\hat{a}^\dagger_k {\rm e}^{-{\rm i}k \cdot x} \right\}.
\end{flalign}
A similar relation holds for the conjugate momentum.}. 
It is crucial to emphasize this point because we will demonstrate that this is no longer the case if the state needs to be further modified by including an exponentiated cubic field operator.

For the modified coherent state \eqref{hom:squeezed}, the expressions of our interest take the following form
\beq
\label{sq:acc}
\partial_t^2\la C_S|\hat{\phi}|C_S\ra(t=0)&=&-\phi_0 \left(m^2+\frac{\lambda}{2}\la S|\hat{\phi}^2|S\ra\right)-\frac{\lambda }{3!}\phi_0^3\,,\\
\la C_S|\hat{H}|C_S \ra &=&\int {\rm d}^3x \left[\frac{1}{2}\left( m^2+\frac{\lambda }{2}\la S|\hat{\phi}^2|S\ra \right)\phi_0^2 +\frac{\lambda }{4!}\phi_0^4 \right.\nonumber \\
&&\hskip 40pt+\frac{1}{2}\la S| \left\{\hat{\pi}^2+ \left(\vec{\nabla}\hat{\phi}\right)^2+m^2\hat{\phi}^2+\frac{\lambda }{4!}\hat\phi^4\right\}|S \ra\nonumber\\
&&\left.\hskip 40pt-\frac{1}{2}\la \Omega| \left\{\hat{\pi}^2+ \left(\vec{\nabla}\hat{\phi}\right)^2+m^2\hat{\phi}^2+\frac{\lambda }{4!}\hat\phi^4\right\}|\Omega \ra\right]\,.
\label{sq:energyS}
\eeq
Here the last two lines of \eqref{sq:energy} originate from the fact that the bare vacuum energy parameter has been adjusted so that $\la\Omega|\hat{H}|\Omega \ra=0$. Moreover, we have assumed, keeping in mind that we are interested in squeezed states, that the expectation value of the odd power of fields in $|S\ra$ vanishes.

Let us begin with the initial acceleration, which at the one-loop level reads
\beq
\label{hom:acc}
\partial_t^2\la C_S|\hat{\phi}|C_S\ra(t=0)=-\phi_0 \left(m^2+\frac{\lambda}{2}\int \frac{{\rm d}^3k}{(2\pi)^3} \frac{1}{2\sqrt{k^2+m^2+\frac{\lambda}{2}\phi_0^2}}\right)-\frac{\lambda }{3!}\phi_0^3\,.
\eeq
The right-hand side immediately strikes us as a Coleman-Weinberg potential. To see explicitly that the standard ($S$-matrix) one-loop renormalization prescription for the quartic theory at hand renders \eqref{hom:acc} finite, let us expand the integrand in the Taylor series in the coupling constant. Up to finite contributions it reduces to
\beq
\frac{\lambda}{2}\int \frac{{\rm d}^3k}{(2\pi)^3} \frac{1}{2\sqrt{k^2+m^2+\frac{\lambda}{2}\phi_0^2}}=\delta m^2+\frac{\delta\lambda}{3!}\phi_0^2+ \text{(finite)}\,,
\eeq
with $\delta m^2$ and $\delta \lambda$ standing for the standard expression of the corresponding counterterms. So that \eqref{hom:acc} repackages into
\beq
\partial_t^2\la C_S|\hat{\phi}|C_S\ra(t=0)=-m_{\rm ph}^2\phi_0 -\frac{\lambda_{\rm ph} }{3!}\phi_0^3+ \text{(finite)}\,,
\eeq
where, as usual, $m_{\rm ph}^2\equiv m^2+\delta m^2$ and $\lambda_{\rm ph}\equiv \lambda+\delta\lambda$. Here (finite) stands for the subtraction-dependent finite contributions.

The procedure outlined above also addresses the one-loop singularities of the Hamiltonian, when applied to equation \eqref{sq:energyS}. 
However, it is crucial to take into account the contributions from the second and third lines. Specifically, in the case of the squeezed vacuum, these lines generate an additional non-trivial correction.
By considering all of these factors together, we find that the renormalization of the potential is achieved in the Hamiltonian by the same squeezing that removes divergences in the equation of motion for the one-point function, as it should.

At this point, it must be clear that we are not bound to choose the squeezed vacuum $|S\ra$ in a way that yields precisely \eqref{modinitphi} and \eqref{modinitpi}. The only parts required by consistency are the divergencies. In other words, we could add any finite contribution to both of those expressions, which would correspond to a specific physical excitation of particles. In fact, we could have chosen to parameterize them as
\beq
\label{modinitphi1}
&&\la S | \hat{\phi}(x,0)\hat{\phi}(y,0) |S\ra=\int \frac{{\rm d}^3k}{(2\pi)^3} \frac{{\rm e}^{{\rm i}\vec{k}\cdot (\vec{x}-\vec{y})}}{2\sqrt{k^2+m^2}}\left(1-\frac{\lambda\phi_0^2}{4}\frac{1}{k^2+m^2} \right)\nonumber\\
&&\hskip 100pt+\int \frac{{\rm d}^3k}{(2\pi)^3} {\rm e}^{{\rm i}\vec{k}\cdot (\vec{x}-\vec{y})}f(k)\,,\\
&&\la S | \hat{\pi}(x,0)\hat{\pi}(y,0) |S\ra=\int \frac{{\rm d}^3k}{(2\pi)^3} ~\frac{1}{2}{\rm e}^{{\rm i}\vec{k}\cdot (\vec{x}-\vec{y})}\sqrt{k^2+m^2}\left( 1+\frac{\lambda\phi_0^2}{4}\frac{1}{k^2+m^2} \right)\nonumber\\
&&\hskip 100pt+\int \frac{{\rm d}^3k}{(2\pi)^3} {\rm e}^{{\rm i}\vec{k}\cdot (\vec{x}-\vec{y})}g(k)\,,
\label{modinitpi1}
\eeq
where $f(k)$ and $g(k)$ are arbitrary functions, regular in the $k\rightarrow 0$ limit and decaying faster than $k^{-3}$ for large momenta. In their absence, the (finite)-terms from the equations given above are also absent. It must be also noted that the magnitudes of these functions control the variance of the state at the initial time and for certain values may result in a classical departure from the validity of the mean-field description of the system.

Notice also that the physical parameters will of course experience running. The imposition of renormalization conditions will fix the value of the potential energy for a certain value of the background field. However, at different values of $\phi_0$, the dynamics and the energy will be determined by the interplay of running and of the initial conditions for the two-point correlation functions.

This is the current state of affairs regarding the expectation values of the field and the Hamiltonian in a scalar field theory with quartic interaction, at one-loop. Next, we delve into the limitations of the squeezing operator when higher loop corrections are taken into account.

\section{Beyond the Squeezing operator and the two-loop dynamics}\label{sec:Tstate}

Now, the crucial question is whether the squeezing alone in the form of \eqref{eq:squeezing_parameter} is capable of eliminating the initial-time singularity across all orders of the loop expansion. This is an important question because semiclassical perturbation theory, when applied to highly occupied systems, typically touches upon initial conditions merely for the mean field and the mode functions for fluctuations. 

Already at the two-loop order, the field renormalization becomes unavoidable. Therefore, we introduce it from the beginning by formulating the theory for physical operators. The Hamiltonian for the renormalized degrees of freedom, which are related to the bare ones by
\beq
\hat{\phi}\rightarrow \sqrt{Z}\hat{\phi}\,, \qquad \hat{\pi}\rightarrow \hat{\pi}/\sqrt{Z}\,,
\eeq
becomes
\beq
\hat{H}=\int {\rm d}^3x\left[ \frac{1}{2Z}\hat{\pi}^2+ \frac{1}{2}Z\left(\vec{\nabla}\hat{\phi} \right)^2+\frac{1}{2}Zm^2\hat{\phi^2}+\frac{\lambda Z^2}{4!}\hat{\phi}^4\right]\,.
\label{HamiltonianZ}
\eeq
Since this redefinition is a canonical transformation, the equal-time commutation relations remain canonically normalized. However, it changes the one for the ladder operators. Taking into account that $\hat{\pi}=Z\partial_t \hat{\phi}$, the following relation follows
\beq
[\hat{\phi}(x),\hat{\pi}(y)]={\rm i}\delta^{(3)}(x-y)\qquad\Longrightarrow \qquad [\hat{a}_k,\hat{a}^\dagger_{k'}]=(2\pi)^3Z^{-1}\delta^{(3)}(k-k')
\label{Zaa}\, .
\eeq

As it was illustrated in~\cite{Berezhiani:2021gph}, the finiteness of the dynamical equation at $t>0$ for the expectation value of the field operator in a non-squeezed coherent state is ensured by the following two-loop mass prescription
\begin{equation}
m^2_\text{ph}=Z m^2+\left(\frac{\lambda}{2}-\frac{\lambda}{2}\int \frac{{\rm d}^3p}{(2\pi)^3(2\omega_p)^3}\right)\langle \hat{\phi^2}\rangle-\frac{\lambda^2}{3}\int\frac{{\rm d}^3p \,{\rm d}^3q}{(2\pi)^6 (8\omega_p\omega_q \omega_{p+q})}\frac{1}{\sum \omega_i},
\label{eq:2loop_mass}
\end{equation}
where $Z$ is the field normalization and $\sum \omega_i=\omega_p+\omega_q+\omega_{p+q}$. In Appendix C we demonstrate that the following prescription is also necessary in order to remove all divergences 
\begin{equation}
    Z=1-\frac{\lambda^2}{3}\int \frac{{\rm d}^3p\,{\rm d}^3q}{(2\pi)^6(8\omega_p\omega_q\omega _{p+q})}\frac{1}{(\sum \omega_i)^3}
\label{eq:2loop_Z}
\end{equation}
(in our previous work~\cite{Berezhiani:2021gph} the divergencies responsible for the field renormalization were not accounted for). It should be emphasized that the divergent parts of integrals appearing in \eqref{eq:2loop_mass} and \eqref{eq:2loop_Z} contain the standard mass and field renormalization divergences, which one encounters in the $S$-matrix analysis at two-loop.

In~\cite{Berezhiani:2021gph}, it was observed that, for non-squeezed coherent states, \eqref{eq:2loop_mass} gives rise to an additional non-renormalizable two-loop initial-time singularity; the same applies to \eqref{eq:2loop_Z}. This is straightforward to see from \eqref{intro:energy}, as non-squeezed coherent states provide merely bubble diagram divergencies at the two-loop order; i.e. the two-loop contribution factorizes into two one-loop integrals.

What about squeezed coherent states? We can begin by examining the additional divergences that a squeezed coherent state with the form \eqref{eq:squeezing_parameter}, which we have invoked for the one-loop renormalization, might introduce at the two-loop level. For starters, let us point out that the introduction of $Z$ factors entails its explicit appearance in the squeezing operator, once written in terms of ladder operators
\beq
    \hat{S}={\frac{1}{2}\int \frac{{\rm d}^3k}{(2\pi)^3}Z\left(\alpha_k \,\hat{a}_k\,\hat{a}_{-k}+\alpha_k^*\,\hat{a}^\dagger_k\,\hat{a}^\dagger_{-k}\right)}\,,
\eeq
which follows from \eqref{squeezeD}. Here $\alpha_k$ has been introduced as a Fourier transform of a generic $D(x-y)$ of \eqref{squeezeD}. Such an appearance allows $Z$ factors to naturally simplify upon the commutation of $\hat{S}$ with various operators.

Specializing for the case $\alpha_k={\rm i}/2\,{\rm ln}(E_k/\omega_k)$ as in \eqref{eq:squeezing_parameter}, it is straightforward to show that only the first divergent two-loop integral of \eqref{eq:2loop_mass} is generated by the squeezing operator. This is accomplished by including corrections that arise from projecting the interacting vacuum of the theory onto the non-interacting one, giving 
\begin{equation}
\label{eq:2pS2loop}
\la S | \hat{\phi}^2(x,0) |S\ra=\int \frac{{\rm d}^3k}{(2\pi)^3 2E_k} \left\{1-\frac{\lambda\langle\phi^2\rangle}{(2\omega_k)^2}\right\}.
\end{equation}
However, notice that this term would have been generated also for non-squeezed coherent states, albeit with $E_k$ replaced by $\omega_k$. This was expected since the bubble diagram is the only divergence that is present also for non-squeezed coherent states.

The second possibility could involve constructing a new squeezing operator designed to produce the missing divergence, represented by the sunset contribution. This can be achieved by choosing a different function $\alpha_k$. However, by implementing this, we would encounter two issues. First, the necessary squeezing parameter we have to introduce must possess a $\phi_0$-independent part. Second, its effect would have to alter quadratic correlation functions in a way that would make the two-point function divergent at point splitting.
Therefore, the squeezing operator cannot, by itself, address the two-loop divergences generated by the sunset contribution to the bare mass and the bare field, which we know for a fact have to be there from the textbook analysis.

The only remaining option for addressing the missing sunset divergence is to go beyond the conventional squeezing and to consider the non-Gaussian extension of squeezed states. In other words, we introduce the following new class of states
\begin{equation}
|C_T\rangle=\exp\left\{{-{\rm i}\int {\rm d}^3 x \, \phi_{0}\,\hat{\pi}(x)}\right\}|T\rangle\qquad \text{with}\quad |T\rangle={\rm e}^{-{\rm i}\hat{S}}e^{-{\rm i}\hat{T}}|\Omega\rangle,
\label{eq:Tstate}
\end{equation}
where $\hat{T}$ is cubic in canonical degrees of freedom, or equivalently in ladder operators. 
We construct it to have the following momentum space representation\footnote{This class of operators in the field space can be written as $\hat{T}=\int {\rm d}^3x \,{\rm d}^3y\, {\rm d}^3z\, M^{\alpha\beta\gamma}(x,y,z)\Phi_\alpha(x) \Phi_\beta(y) \Phi_\gamma (z)$, with $\Phi_1=\hat{\phi}$ and $\Phi_2=\hat{\pi}$.}
\beq
\hat{T}=\frac{1}{3}\int \frac{{\rm d}^3k\, {\rm d}^3 k'}{(2\pi)^6} \left\{\gamma^*_{k,k'}\hat{a}^\dagger_k \hat{a}^\dagger_{k
    '}\hat{a}^\dagger_{-k-k'}+\gamma_{k,k'}\hat{a}_k \hat{a}_{k
    '}\hat{a}_{-k-k'}\right\}\,,
    \label{Tgamma}
\eeq
with $\gamma_{k,k'}$ being a complex function of the momenta, to be specified later. We have chosen not to introduce operators of the mixed form $\hat{a}_k^\dagger \hat{a}_{k'} \hat{a}_{-k-k'}$, since doing so would alter the initial conditions of the one-point function of the system. The ordering of the squeezing operator and $\hat{T}$ within the state is also not arbitrary. The adopted order ensures that when a correlation function is computed on the state given by \eqref{eq:Tstate}, the dispersion relations resulting from the Fourier decomposition of fields correspond to the dispersion with a shifted mass, owing to the influence of $\hat{S}$. Not to mention that all quantum fields undergo a shift at the initial time, due to the field displacement induced by \eqref{eq:Tstate}.
 In other words,  at the initial time a general equal-time correlation function evaluated on \eqref{eq:Tstate} takes the following form 
\begin{equation}
    \langle C_T|\mathcal{O}(\hat{\phi},\hat{\pi})|C_T\rangle=\langle \Omega|\mathcal{O}(\phi_0+{\rm e}^{-{\rm i}\hat{T}}\hat{\phi}_s{\rm e}^{{\rm i}\hat{T}},{\rm e}^{-{\rm i}\hat{T}}\hat{\pi}_s{\rm e}^{{\rm i}\hat{T}})|\Omega\rangle\,.
\end{equation}

Interestingly, unlike the squeezing and the shift operator, the action of $\hat{T}$ upon operators does not appear to have a resummed form. It is straightforward to see that the unitary operator ${\rm e}^{{\rm i}\hat{T}}$ is not a rotation in the space of ladder operators, due to its nonlinear nature, even with respect to a generic operator as a parameter of a Bogoliubov transformation. Namely, one readily obtains that
\beq
&&{\rm e}^{{\rm i}\hat{T}}\hat{a}_p \,{\rm e}^{-{\rm i}\hat{T}}=\hat{a}_p-\int \frac{{\rm d}^3k_1}{(2\pi)^3} \, \gamma_{p,k_1}\,\hat{a}^\dagger_{k_1}\hat{a}^\dagger_{-k_1-p}\nonumber\\
&&\hskip 70pt+\int \frac{{\rm d}^3k_1 \,{\rm d}^3k_2}{(2\pi)^6}\,\gamma_{p,k_1}\gamma_{k_1,k_2}\left(\hat{a}_{k_2}\hat{a}_{k_1}\hat{a}^\dagger_{k_1+k_2-p}+\hat{a}^\dagger_{k_1+k_2-p} \hat{a}_{k_1}\hat{a}_{k_2}\right)+...\,.
     \label{eq:Ta}
\eeq
Clearly, the result of a non-Gaussian extension of the coherent state is beyond the Bogoliubov transformation due to the appearance of nested momentum integrals. Moreover, the series does not seem to resum in any particular form, rather the operators of increasing complexity seem to emerge.

\subsection{Removing the two-loop initial-time singularity}


Let us now showcase an explicit form of $|T\rangle$ which generates additional divergences at the initial time, required by the two-loop mass and field renormalization. For this, we merely need up to $\mathcal{O}(\lambda^2)$ contribution to the equation of motion and the Hamiltonian. The extension of the analysis to account for two-loop coupling constant renormalization would inevitably invoke $\mathcal{O}(\lambda^3)$-terms which would make the presentation cumbersome. On the other hand, we believe that the demonstration with merely mass and field renormalization serves as a sufficient proof of principle.

A straightforward analysis shows that an adequate choice for $\gamma_{k,k'}$ coefficients of \eqref{Tgamma} fulfilling our needs can be reduced to
\begin{equation}
\hat{T}={\rm i}\frac{\lambda}{6}\int \frac{{\rm d}^3k \,{\rm d}^3 k'}{(2\pi)^6\sqrt{8E_kE_{k'}E_{k+k'}}} \frac{\hat{a}_k \hat{a}_{k
    '}\hat{a}_{-k-k'}-\hat{a}^\dagger_k \hat{a}^\dagger_{k
    '}\hat{a}^\dagger_{-k-k'}}{\sum E_i}\left(\phi_0-\frac{\ddot{\phi}_0}{(\sum E_i)^2}\right),
\label{eq:Tsqueezing}
\end{equation}
with $\sum E=E_k+E_{k'}+E_{k+k'}$ and $\ddot{\phi}_0$ denoting the initial classical field acceleration. For the purposes of examining $\mathcal{O}(\lambda^2)$ terms, we have $\ddot{\phi}_0=-m^2\phi_0$.
The form of \eqref{eq:Tsqueezing} has been designed in order to eliminate the initial time singularity arising from two-loop mass and field renormalization. 
Notice that, as long as we do not extend the analysis to two-loop coupling constant counterterms, we can use  $E_k$ and $\omega_k$ interchangeably in \eqref{eq:Tsqueezing}.  Moreover, the field renormalization factor $Z$ has been set to unity within $\hat{T}$, as the resulting error is beyond the loop order of our interest here.

Notice that momentum-dependent coefficients other than \eqref{eq:Tsqueezing} can be considered as well. The key point is that, similar to the squeezing operators, we need to ensure the presence of a certain minimal non-Gaussianity in an interacting field theory. Nevertheless, we retain the freedom to have nontrivial finite momentum imprints.

For our discussion, it suffices to compute the leading corrections induced on the two and three-point correlation functions (see Appendix B for details)
\begin{flalign}
\label{eq:2pT}
&\la T | \hat{\phi}^2 |T\ra_0=\int \frac{{\rm d}^3 k}{(2\pi)^3 2E_k} \left\{1-\frac{\lambda\langle\phi^2\rangle}{(2\omega_k)^2}+\lambda^2\int\frac{{\rm d}^3\,k'}{(2\pi)^38E_k E_{k'}E_{k+k'}}\left(\frac{\phi_0}{\sum E_i}-\frac{\ddot{\phi}_0}{(\sum E_i)^3}\right)^2\right\}\,,\\[0.18cm]
&\la T | \hat{\phi}^3|T\ra_0=-{\lambda }\int \frac{{\rm d}^3k\,{\rm d}^3k'}{(2\pi)^6{4E_k E_{k'}E_{k+k'}}}\frac{1}{\sum E_i}\left(\phi_0-\frac{\ddot{\phi}_0}{(\sum E_i)^3}\right)\,.
\label{eq:3pT}
\end{flalign}
Here, correlation functions are given at coincidence to keep their expression compact. One can straightforwardly verify that the modified correlation functions are finite at point-splitting. Higher order corrections induced by $|T\rangle$ would contribute starting with three-loops.
It must be noted that the inclusion of $\hat{T}$ introduces non-trivial conditions for the connected part of the initial 3-point function, which was previously absent.

We proceed by examining expectation values of the equation of motion and the Hamiltonian, as in the one-loop case. The former, evaluated on the state $|C_T\rangle$ at the initial moment of time, reduces to
\beq
\label{eq:1pT}
Z\partial_t^2\la C_T|\hat{\phi}|C_T\ra(t=0)=-\phi_0 \left(Z m^2+\frac{\lambda}{2}\la T | \hat{\phi}^2(x,0)|T\ra\right)-\frac{\lambda Z^2}{3!}\phi_0^3-\frac{\lambda}{3!}\la T | \hat{\phi}^3(x,0)|T\ra\,.
\label{eq:EOMopT}
\eeq
Since the non-trivial part of $Z$ starts at order $\hbar^2$, we introduced it only in front of terms which have a classical part.

By expanding the quadratic and cubic correlation functions \eqref{eq:2pT} and \eqref{eq:3pT} in $\phi_0$ and substituting them into the equation of motion \eqref{eq:1pT}, we obtain
\begin{flalign}
&Z\partial_t^2\la C_T|\hat{\phi}|C_T\ra(t=0)=\nonumber-\ddot{\phi}_0\left(\frac{\lambda^2}{3}\int \frac{{\rm d}^3k\,{\rm d}^3k'}{(2\pi)^6{8\omega_k \omega_{k'}\omega_{k+k'}}}\frac{1}{\left(\omega_k+\omega_{k'}+\omega_{k+k'}\right)^3}\right)\\&-\phi_0\biggr(Z m^2+ \left(\frac{\lambda}{2}-\frac{\lambda^2}{2}\int \frac{{\rm d}^3k}{(2\pi)^3 (2\omega_k)^3}\right)\langle\phi^2\rangle-\frac{\lambda^2}{3}\int\frac{{\rm d}^3k \,{\rm d}^3k'}{(2\pi)^68 \omega_k \omega_{k'} \omega_{k+k'}}\frac{1}{\omega_{k}+\omega_{k'}+\omega_{k+k'}}\biggr)\nonumber\\&+\frac{\lambda^{\rm 1-loop}_{\rm ph}}{3!}\phi_0^3+\mathcal{O}(\lambda^3)\,.
\end{flalign}
Notice that the first $\lambda^2$ term of the second line is the one obtained by projecting the interacting vacuum of the theory on the non-interacting vacuum.
The expression of the second line inside the parentheses is precisely the two-loop $m^2_\text{ph}$ defined in \eqref{eq:2loop_mass}. The first term on the right-hand side (i.e. the first line) combines with the left-hand side to cancel its divergent part exactly. This is precisely what we designed $\hat{T}$ in \eqref{eq:Tsqueezing} for, hence the result should not be surprising. 

Finally, we turn to the expectation value of the Hamiltonian, and verify its finiteness. In this case, we want to focus on the $\lambda^2$  divergent terms appearing in 
\beq
\la C_T|\hat{H}|C_T \ra &=&\int {\rm d}^3x \left[\frac{Z}{2}\left( m^2+\frac{\lambda }{2}\la T|\hat{\phi}^2|T\ra \right)\phi_0^2 +\frac{\lambda Z^2 }{4!}\phi_0^4+\frac{\lambda }{3!}\langle T|\hat{\phi}^3|T\rangle \phi_0+ \right.\nonumber \\
&&\hskip 40pt+\frac{1}{2}\la T| \left\{\hat{\pi}^2+ \left(\vec{\nabla}\hat{\phi}\right)^2+m^2\hat{\phi}^2+\frac{\lambda }{4!}\hat\phi^4\right\}|T \ra\nonumber\\
&&\left.\hskip 40pt-\frac{1}{2}\la \Omega| \left\{\hat{\pi}^2+ \left(\vec{\nabla}\hat{\phi}\right)^2+m^2\hat{\phi}^2+\frac{\lambda }{4!}\hat\phi^4\right\}|\Omega \ra\right]\,.
\label{sq:energy}
\eeq

At first glance, the divergence from the cubic correlator in \eqref{sq:energy} does not combine with the bare mass as nicely as in the equation of motion due to the difference in coefficients. A similar situation emerges for the term proportional to $\ddot{\phi}_0$ due to the absence of a classical kinetic term for the initial configuration depicted by the state \eqref{eq:Tstate}.

However, in this case, quadratic correlators are also important and we can check that the following relation holds for the state
\begin{flalign}
  \frac{1}{2} \la T| \hat{\pi}^2+ \left(\vec{\nabla}\hat{\phi}\right)^2+&m^2\hat{\phi}^2|T\ra=\frac{1}{2} \la S| \hat{\pi}^2+ \left(\vec{\nabla}\hat{\phi}\right)^2+m^2\hat{\phi}^2|S\ra\nonumber\\+&\frac{\lambda^2 \phi_0}{3!2} \int\frac{{\rm d}^3k'\,{\rm d}^3k}{(2\pi)^6 8E_k E_{k'}E_{k+k'}}\frac{1}{\sum E_i}\left(\phi_0-\frac{2 \ddot{\phi}_0 }{(\sum E_i)^2}\right)\ +\ \text{(finite)}\ +\ \mathcal{O}(\lambda^3)\,.
  \label{eq:2looprelation}
\end{flalign}
This can be derived by explicitly evaluating the combination of operators present on the left-hand side and comparing it to the right-hand side.

Using \eqref{eq:2looprelation} back in \eqref{sq:energy},
we see that the term proportional to $\phi^2_0$ combines with analogous contributions from the cubic correlation function in such away that the overall divergence is eliminated via the mass prescription \eqref{eq:2loop_mass}. In contrast, the two terms proportional to $\phi_0\ddot{\phi}_0$ sum up to zero. This makes the expectation value of the Hamiltonian finite.

\section{Quantum Electrodynamics}
\label{sec:gauge}

We would like to finish the discussion of the initial-time singularities by examining coherent states within electrodynamics coupled to scalar charges. In particular, we would like to examine the perturbativity of coherent states within the framework of the so-called BRST invariant quantization, as implemented in~\cite{Berezhiani:2021zst}. We will not go into details of the framework here, for the basics of the formalism we refer the reader to~\cite{Kugo:1977zq,Kugo:1977yx,Kugo:1977mk,Kugo:1977mm}; see also~\cite{Weinberg:1996kr}.

The starting point is the gauge-fixed Lagrangian supplemented with Faddeev-Popov ghosts
\beq
\label{enmlag}
\mathcal{L}=-\frac14 \hat{F}_{\mu\nu}^2+|D_\mu\hat{\Phi}|^2-m^2|\hat{\Phi}|^2-\partial_\mu \hat{b}\hskip 1pt \hat{A}^\mu+\frac12\xi \hat{b}^2+\partial_\mu \hat{\bar c}\hskip 1pt \partial^\mu \hat{c}\,,
\eeq
with $\hat{F}\mn\equiv\partial_\mu \hat{A}_\nu-\partial_\nu \hat{A}_\mu$ and $D_\mu\hat{\Phi}\equiv\partial_\mu\hat{\Phi}-ig\hat{A}_\mu\hat{\Phi}$. Here the gauge fixing has been implemented with the aid of the auxiliary degree of freedom $\hat{b}$, which becomes a canonical conjugate of $\hat{A}_0$. The Faddeev-Popov fields $\hat{c}$ and $\hat{\bar{c}}$ are anti-commuting Lorentz scalars, as usual. Such a formulation automatically fixes the issue of gauge redundancy and the subsequent difficulties with the quantization. At first sight, this construction seems to come with a price of propagating unphysical degrees of freedom, namely the longitudinal photon together with a temporal component and ghost fields. However, at the same time this procedure gives rise to the celebrated BRST symmetry, emerging as a residual symmetry of the adopted gauge fixing procedure. The associated fermionic conserved charge $\hat{Q}$ serves as an important discriminator in constructing the physical Hilbert space of the theory. In particular, physical states in this framework are those that are annihilated by the BRST charge, $\hat{Q}|\text{phys}\ra=0$.

The physical, i.e., BRST invariant, coherent states for the electromagnetic field can be constructed as~\cite{Berezhiani:2021zst}
\beq
\label{glauberst}
    |A\rangle={\rm e}^{-{\rm i}\int {\rm d}^3x \left(A^c_j(x)\hat{E}_j-E_j^c(x)\hat{A}_j+A_0^c\hat{b}\right)}|\Omega\rangle\,,
\eeq
where the quantities with a label '$c$' are the c-number functions parametrizing the state, with $E_j^c$ satisfying the sourceless Gauss' law $\partial_j E_j^c=0$. These functions set the initial one-point function of the corresponding operators. This class of states was shown to be annihilated by $\hat{Q}$, for arbitrary $A_\mu^c$ due to the BRST invariance of $\hat{E}_j$ and $\hat{b}$. It is important to point out that there is a subclass of these c-number functions corresponding to the pure-gauge configurations, so that the corresponding \eqref{glauberst} are physically equivalent (at the $S$-matrix level) to the vacuum state with vanishing vector potential~\cite{Berezhiani:2021zst,LGO}.

These are the so-called Glauber-Sudarshan states within the BRST formalism. They are non-squeezed coherent states since they are characterized merely by the one-point expectation values of the electromagnetic fields. The goal of this section is to demonstrate that such states seem to exhibit similar perturbative problems with renormalizability once they are introduced within the interacting theory \eqref{enmlag}, in a similar way as the scalar field example considered in previous sections.

To show that \eqref{glauberst} is indeed in need of squeezing, it is instructive to examine the expectation value of the Hamiltonian that corresponds to \eqref{enmlag}. As shown in~\cite{Berezhiani:2021zst},  this is straightforward to obtain and is given by
\beq
\label{hamilton}
\hat{H}=\int {\rm d}^3 x\left[ \frac{1}{2}\hat{E}_j^2+\frac{1}{4}\hat{F}_{ij}^2+|\hat{\Pi}|^2+|\hat{D}_j\hat{\Phi}|^2+m^2|\hat{\Phi}|^2
-\hat{A}_j\p_j\hat{b}-\frac{\xi}{2}\hat{b}^2 \right.+\nonumber\\
+  \hat{E}_j\p_j \hat{A}_0+{\rm i}g\hat{A}_0\left( \hat{\Phi}\hat{\Pi}-\hat{\Pi}^\dagger\hat{\Phi}^\dagger \right)
+\hat{\Pi}_c\hat{\Pi}_{\bar c}+\p_j \hat{\bar{c}}~\p_j\hat{c}+\Lambda_{\rm B}-g\hat{A}_0 \rho_{\rm vac}\Big]\,.~
\eeq
Here $\Lambda_{\rm B}$ denotes the bare vacuum energy that we adjust in such a way that $\la \Omega |\hat{H}| \Omega\ra=0$. Moreover, in the last term $\rho_{\rm vac}$ refers to an infinite vacuum charge density. It has been introduced to remove the latter from the Gauss' law, i.e. $\rho_{\rm vac}\equiv {\rm i}\la\Omega| \hat{\Phi}\hat{\Pi}-\hat{\Pi}^\dagger\hat{\Phi}^\dagger |\Omega\ra$. Naively, one could associate the introduction of such a counter-term with the Lorentz violation. However, in this case, it remains innocuous as it does not give rise to any physical effects, apart from cancelling the vacuum charge divergence.  As a result, the expectation value of the Hamiltonian in \eqref{glauberst} reduces to
\beq
\la A | \hat{H} | A \ra = \int {\rm d}^3x\bigg[\frac{1}{2}\left( E_c^2+B_c^2 \right)+g^2\left(A_j^c\right)^2\la \Omega | \hat{\Phi}^\dagger\hat{\Phi}|\Omega \ra\bigg]-\int {\rm d}S_j\left(E_j^cA_0^c\right)\,,
\label{qedHvev}
\eeq
where $B_c$ is the magnetic field, with $B_i^c\equiv \epsilon_{ijk}\p_j A_k^c$. We have also taken into account that the vacuum current vanishes, i.e. ${\rm i}\la \Omega | \hat{\Phi}\p_j \hat{\Phi}^\dagger -\hat{\Phi}^\dagger \p_j \hat{\Phi} |\Omega \ra=0$. The last term of \eqref{qedHvev} is the surface integral, which can be dismissed for configurations that die off fast enough at the boundary.

Evidently, we are encountering the same perturbative issues here as we did for the non-squeezed coherent states of the scalar field theory. We can begin by noticing that there is a background-dependent divergent contribution in \eqref{qedHvev}, and no bare parameters to hide it with. However, there is one infinite constant we have not manifested yet. This is the field renormalization constant which has to be infinite as we learn from the $S$-matrix analysis. Given that we would aim to initialize the renormalized electromagnetic field as $E_c$ and $B_c$, the field normalization that would ensure this corresponds to introducing $Z$ factor by replacing $E_c\rightarrow Z^{-1/2}E_c$ and $B_c\rightarrow Z^{1/2}B_c$\label{eq:GaugeH}. However, it is evident that this does not solve the situation: the terms proportional to $Z$ and the term proportional to the scalar field bubble have a different background dependence and cannot be combined. In fact, the introduction of $Z$-factors introduces additional divergences that need to be addressed in this context.

All this seems to point towards the necessity for going beyond the simple non-squeezed coherent states in quantum electrodynamics as well.
In particular, the squeezing of scalar modes seems to be a necessary condition to make the energy \eqref{eq:GaugeH} finite. This could have interesting consequences, as it would imply that the initial conditions for a spectator field on an electromagnetic background cannot be arbitrarily chosen.

\section{Summary}
\label{sec:summary}

In this work, we have examined the perturbative construction of coherent states as a fundamental description of classical backgrounds. The issue we have set out to resolve is best elucidated by beginning with a theory of a free massive scalar field in the Schr\"odinger picture, governed by the Hamiltonian
\beq
\hat{H}_0=\int {\rm d}^3 x \left( \frac{1}{2}\hat{\pi}^2+\frac{1}{2}\left(\partial_j\hat{\phi}\right)^2+\frac{1}{2}m^2\hat{\phi}^2+\Lambda \right)\,.
\eeq
Here $\Lambda$ stands for the bare vacuum energy that we use to adjust the lowest energy eigenvalue to zero. The corresponding (vacuum) eigenstate is denoted by $|0\ra$. In this case the coherent state corresponding to a homogeneous field configuration can be straightforwardly constructed as
\beq
|C\ra (t=0)={\rm e}^{-{\rm i}\int{\rm d}^3 x\,\phi_0\,\hat{\pi}}|0\ra\,.
\label{freecohst}
\eeq
Even the time evolution would be straightforward to compute and would give
\beq
|C\ra (t)={\rm e}^{-{\rm i}\hat{H}_0 t}\,|C\ra (t=0)={\rm e}^{-{\rm i}\int {\rm d}^3 x\,\phi_0\left({\rm cos}(mt)\hat{\pi}+m\hskip 1pt {\rm sin}(mt)\hat{\phi}\right)}|0\ra\,,
\eeq
which is precisely what is expected from the classical dynamics.

The question is how to modify \eqref{freecohst} consistently upon introducing an interaction term. For this, one could start by considering the interacting theory
\beq
\hat{H}=\hat{H}_0+\int {\rm d}^3x\left(\frac{\lambda}{4!}\hat{\phi}^4\right)\,,
\label{fullH}
\eeq
together with a readjustiment of $\Lambda$, to make the lowest energy eigenstate $|\Omega\ra$ to have a vanishing eigenvalue, i.e. $\hat{H}|\Omega\ra=0$.
One could then guess that a natural progression to modify \eqref{freecohst} for this interacting theory is to replace
\beq
|0\ra\quad\longrightarrow \quad|\Omega\ra\,.
\eeq
In particular, as we have shown in~\cite{Berezhiani:2020pbv,Berezhiani:2021gph}, this adjustment generates vacuum bubble contributions to correlation functions, which otherwise would be missing. This was also pointed out in the String Theory context in~\cite{Brahma:2020tak}.

From a first look, the state resulting from this choice
\beq
|C\ra (t=0)={\rm e}^{-{\rm i}\int {\rm d}^3 x\, \phi_0\,\hat{\pi}}|\Omega\ra\,
\label{intcohst}
\eeq
seems to have all the right ingredients. It is constructed from the vacuum of the theory and canonical degrees of freedom, in this case the conjugate momentum. Notice that in the Heisenberg picture, the operators at hand would have been canonical degrees of freedom of the full theory with the time evolution dictated by \eqref{fullH}, rather than the ones of the interaction picture with a free-theory dynamics.
This state is a representative of a more general class of states given by \eqref{intro:coherent}, which we refer to as non-squeezed coherent states. They seem interesting as they are elegant to manipulate, for being generated from the vacuum by merely a field-displacement operator.

However, as we demonstrated in~\cite{Berezhiani:2020pbv,Berezhiani:2021gph} and have reiterated here as well, such states lead to divergent contributions to physical quantities in perturbation theory. For example, the expectation value of the Hamiltonian in \eqref{intcohst} exhibits an unrenormalizable feature already at one-loop order. We have shown that to fix this issue it is necessary to expand the discussion into the realm of squeezed coherent states. The modification one needs to invoke corresponds to altering initial conditions for the two-point function of \eqref{intcohst}. In fact, the required initial state is such that it corresponds to the vacuum of the Bogoliubov modes for fluctuations of momenta $k\gg \sqrt{\lambda} \phi_0$. This was noticed for the first time in~\cite{Baacke:1997zz} in the context of the semiclassical analysis and we highlighted it in the context of coherent states.

An important follow-up point we discussed in this work is the failure of squeezed coherent states in addressing initial time singularities, once two-loop corrections are included.
The extension of the analysis to this order revealed that the correct construction of coherent states in interacting theories requires the replacement 
\begin{equation}
|\Omega\rangle\quad \longrightarrow \quad |T\rangle,
\end{equation}
where $|T\rangle$ is the non-gaussian vacuum, built by acting on the true vacuum with exponential operators, with a \textit{cubic} exponent in the field operator and its conjugate momentum. In particular, the modification of the initial-time cubic correlation functions is crucial to reproduce all two-loop divergences in the initial time field acceleration and the energy. Although these corrections disappear once the classical limit of the theory is considered,
they imply that coherent (squeezed) states do not have good perturbative behaviour when quantum corrections are included.

\section*{Acknowledgements}

We would like to thank Gia Dvali and Otari Sakhelashvili for invaluable discussions.

\appendix

\section{Correlation  functions evaluated over the Squeezed state}
\renewcommand{\theequation}{A-\Roman{equation}}
\setcounter{equation}{0}

In this appendix, we show how to evaluate correlation functions over the squeezed state $|S\rangle={\rm e}^{-{\rm i}\hat{S}}|\Omega\rangle$.  We assume a hermitian squeezing operator $\hat{S}=\hat{S}^\dagger$, which we parameterize as
\begin{equation}
    \hat{S}={\frac{1}{2}\int \frac{{\rm d}^3k}{(2\pi)^3}\left(\alpha_k \,\hat{a}_k\,\hat{a}_{-k}+\alpha_k^*\,\hat{a}^\dagger_k\,\hat{a}^\dagger_{-k}\right)},
\end{equation}
where $\alpha_k$ is a c-number function.

To evaluate a general correlation function over this state, we start by applying the following notable relation
\begin{equation}
    \langle S|\mathcal{O}\big(\hat{\phi},\hat{\pi}\big)|S\rangle= \langle \Omega| {\rm e}^{{\rm i}\hat{S}}\mathcal{O}\big(\hat{\phi},\hat{\pi}\big){\rm e}^{-{\rm i}\hat{S}}|\Omega \rangle=\langle \Omega| \mathcal{O}\big({\rm e}^{{\rm i}\hat{S}}\hat{\phi}\,{\rm e}^{-{\rm i}\hat{S}},{\rm e}^{{\rm i}\hat{S}}\hat{\pi}{\rm e}^{-{\rm i}\hat{S}}\big)|\Omega \rangle,
\label{eq:formulaUnitary}
\end{equation}
with $\mathcal{O}\big(\hat{\phi},\hat{\pi}\big)$ an arbitrary function of $\hat{\phi}$ and $\hat{\pi}$.
To derive the last step, we expand $\mathcal{O}$ in powers of the field operator and then we insert the identity operator ${\rm \hat{I}} ={\rm e}^{i\hat{S}}{\rm e}^{-i\hat{S}}$ in between each power of $\hat{\phi}$ and $\hat{\pi}$. Finally, we resum the series.

We read from Eq.~\eqref{eq:formulaUnitary} that any correlation function evaluated over $|S\rangle$ can be equivalently computed as a vacuum correlation function after making the following replacement
\begin{flalign}
\hat{\phi}\rightarrow {\rm e}^{{\rm i}\hat{S}} \hat{\phi}\, {\rm e}^{-{\rm i}\hat{S}}\label{eq:phis},  \\
\hat{\pi}\rightarrow {\rm e}^{{\rm i}\hat{S}} \hat{\pi}\, {\rm e}^{-{\rm i}\hat{S}}\label{eq:pis}.
\end{flalign}
 Relations \eqref{eq:phis} and \eqref{eq:pis} can be expanded by applying the Baker–Campbell–Hausdorff formula as
\begin{equation}
    {\rm e}^{{\rm i}\hat{S}}\,\hat{\phi}\, {\rm e}^{-{\rm i}\hat{S}}=\phi+[{\rm i}\hat{S},\hat{\phi}]+\frac{1}{2}[{\rm i}\hat{S},[{\rm i}\hat{S},\hat{\phi}]]+\frac{1}{3!}[{\rm i}\hat{S},[{\rm i}\hat{S},[{\rm i}\hat{S},\hat{\phi}]]]+...,\label{eq:UBCH}
\end{equation}
We use the explicit form of $\hat{S}$ to evaluate the series of commutators \eqref{eq:UBCH}. 
The series of commutators can be resummed.
By writing $\alpha_k=|\alpha_k|{\rm e}^{{\rm i}\theta_k}$, we find the closed expression
\begin{flalign}
\hat{\phi}_s=\int&\frac{{\rm d}^3 k}{(2\pi)^3\sqrt{2\omega_k}}\biggr\{\left(\cosh |\alpha_k|+{\rm i}\sinh|\alpha_k|{\rm e}^{{\rm i}\theta_k}\right) \hat{a}_k +\left(\cosh |\alpha_k|-{\rm i}\sinh|\alpha_k|{\rm e}^{-{\rm i}\theta_k}\right) \hat{a}^\dagger_{-k} \biggr\}{\rm e}^{{\rm i}k \cdot x} ,
\end{flalign}
which is a Bogoliubov transformation of parameters $|\alpha_k|$ and $\theta_k$, acting on the original $a_k$ and $a^\dagger_k$. Here, $\omega_k=\sqrt{k^2+m^2}$. A similar relation holds for the conjugate momentum
\begin{flalign}
\hat{\pi}_s=(-{\rm i})\int\frac{{\rm d}^3 k}{(2\pi)^3}\sqrt{\frac{\omega_k}{2}}\biggr\{\left(\cosh |\alpha_k|-{\rm i}\sinh|\alpha_k|{\rm e}^{{\rm i}\theta_k}\right) \hat{a}_k -\left(\cosh |\alpha_k|+{\rm i}\sinh|\alpha_k|{\rm e}^{-{\rm i}\theta_k}\right) \hat{a}^\dagger_{-k} \biggr\}{\rm e}^{{\rm i}k \cdot x}
\end{flalign}
By introducing the parameters  $\theta_k=\frac{\pi}{2}$ and $|\alpha_k|=\frac{1}{2}\ln\left(\frac{E_k}{\omega_k}\right)$, with $E^2_k=k^2+m^2+\frac{\lambda}{2}\phi_0^2$, we obtain
\begin{flalign}
 & \label{rotphi2}\hat{\phi}_s=\int\frac{{\rm d}^3 k}{(2\pi)^3\sqrt{2E_k}}\left( \hat{a}_k  {\rm e}^{{\rm i}k \cdot x}+ \hat{a}^\dagger_k {\rm e}^{-{\rm i}k \cdot x} \right),\\
     &\hat{\pi}_s=(-{\rm i})\int\frac{{\rm d}^3 k}{(2\pi)^3}\sqrt{\frac{E_k}{2}}\left( \hat{a}_k  {\rm e}^{{\rm i}k \cdot x}- \hat{a}^\dagger_k {\rm e}^{-{\rm i}k \cdot x} \right).
     \label{rotpi2}
\end{flalign}
Correlation functions \eqref{modinitphi} and \eqref{modinitpi} at the initial time follow from this new operators \eqref{rotphi2} and \eqref{rotpi2}. 

\section{Correlation  functions evaluated over the \textit{T}-state}
\renewcommand{\theequation}{B-\Roman{equation}}
\setcounter{equation}{0}

In this appendix, we derive the procedure to evaluate correlation functions over the non-Gaussian state
\begin{equation}
    |T\rangle={\rm e}^{{\rm i} \hat{T}}|\Omega\rangle,\quad\text{with}\quad \hat{T}=\frac{1}{3}\int \frac{{\rm d}^3k {\rm d}^3 k'}{(2\pi)^6} \left\{\gamma^*_{k,k'}\hat{a}^\dagger_k \hat{a}^\dagger_{k
    '}\hat{a}^\dagger_{-k-k'}+\gamma_{k,k'}\hat{a}_k \hat{a}_{k
    '}\hat{a}_{-k-k'}\right\}.
\end{equation}
Since $\hat{T}$ is hermitian, we can apply the formula \eqref{eq:formulaUnitary}, albeit with $\hat{S}$ replaced by $\hat{T}$. In other word, we have
\begin{equation}
    \langle T|\mathcal{O}\big(\hat{\phi},\hat{\pi}\big)|T\rangle=\langle \Omega| \mathcal{O}\big({\rm e}^{{\rm i}\hat{T}}\hat{\phi}\,{\rm e}^{-{\rm i}\hat{T}},{\rm e}^{{\rm i}\hat{T}}\hat{\pi}{\rm e}^{-{\rm i}\hat{T}}\big)|\Omega \rangle.
\label{eq:formulaUnitary2}
\end{equation}

Now, we are left with evaluating the $T$-rotated field operators, by applying the analog of \eqref{eq:UBCH}.
In this case, there is no closed form for the series of commutators since the number of creation and annihilation operators increases as we increase the number of nested commutators. 
However, since we are interested in the effect of $\hat{T}$ up to $\hbar^2$ corrections, we can evaluate the series of commutators up to
\begin{equation}
\hat{\phi}_T={\rm e}^{{\rm i}\hat{T}}\,\hat{\phi} \,{\rm e}^{{\rm i}\hat{T}} \simeq  \hat{\phi}+[{\rm i}\hat{T},\phi]+\frac{1}{2}[{\rm i}\hat{T},[{\rm i} \hat{T},\hat{\phi}]]+...
\end{equation}
By evaluating the commutators explicitly, we find that $\hat{T}$ acts on the field operator with the non-trivial transformation
\begin{flalign}
\hat{\phi}\,\rightarrow &\,\hat{\phi}_T=\hat{\phi}+{\rm i}\int \frac{{\rm d}^3k\,{\rm d}^3k'}{\left(2\pi\right)^6\sqrt{2\omega_k}}\left(\gamma_{k,k'}\hat{a}_{k'}\hat{a}_{-k-k'}{\rm e}^{-{\rm i}k\cdot x}-\gamma^*_{k,k'}{\rm e}^{{\rm i}k\cdot x}\hat{a}^\dagger_{k'}\hat{a}^\dagger_{-k-k'}\right)\nonumber\\&+\frac{1}{3}\int\frac{{\rm d}^3k\,{\rm d}^3k'\,{\rm d}^3k''}{(2\pi)^6\sqrt{2\omega_k}}\gamma_{k,k'}\gamma^*_{k,k''}{\rm e}^{{\rm i}k\cdot x}\left(\hat{a}^\dagger_{k''}\hat{a}_{k'}\hat{a}_{-K}+\hat{a}_{k'}\hat{a}^\dagger_{k''}\hat{a}_{-K}+\hat{a}_{k'}\hat{a}_{-K}\hat{a}^\dagger_{k''}\right)+\text{h.c},
\end{flalign}
where $K=k+k'+k''$.
Here, h.c. is the hermitian conjugate of the second line.
Also, we are assuming that $\gamma_{k_1,k_2}$ is invariant under the momentum redefinition $k_1\rightarrow -k_1-k_2$ and $k_2\rightarrow -k_1-k_2$. 

Using this redefinition, we can compute the quadratic and cubic correlation functions at the initial-time
\begin{flalign}
    \langle T|\hat{\phi}^3(x)|T\rangle=\langle \Omega|\hat{\phi}_T^3(x)|\Omega\rangle={\rm i}\int \frac{{\rm d}^3k \,{\rm d}^3k'}{(2\pi)^6\sqrt{2 \omega_k\omega_{k'}\omega_{k+k'}}}\left(\gamma_{k,k'}-\gamma^*_{k,k'}\right)\\
      \langle T|\hat{\phi}^2(x)|T\rangle=\langle \Omega|\hat{\phi}_T^2(x)|\Omega\rangle=
    \int\frac{{\rm d}^3k}{(2\pi)^3 2\omega_k}\left(1+\int \frac{ {\rm d}^3k'}{(2\pi)^3 }4|\gamma_{k,k'}|^2\right).
\end{flalign}
By choosing the $\gamma_{k,k'}$ coefficient accordingly, the quadratic and cubic correlation functions \eqref{eq:2pT} and \eqref{eq:3pT} are reproduced.

\section{Renormalizing the Equation of Motion for the One-Point Function of the Field Operator}
\renewcommand{\theequation}{C-\Roman{equation}}
\setcounter{equation}{0}

In this appendix, we prove that the two-loop field prescription 
\eqref{eq:2loop_Z} makes the one-point function of the field operator finite at $t>0$.

Let us start by recalling a result from~\cite{Berezhiani:2021gph}. The equation of motion for $\Phi$, evaluated up to $\lambda^2$ and $\hbar^2$ corrections, reads
\begin{flalign}
Z\nonumber\left(\partial_t^2+m^2\right)&\Phi(t)+\frac{Z^2\lambda}{3!}\Phi^3(t)+\frac{\lambda}{2}\langle\phi^2\rangle\Phi(t)-\frac{\lambda^2}{2}\Phi(t)\int \frac{{\rm d}^3p}{(2\pi)^3(2\omega_p)^2}\int_0^t{\rm d}t_1\Phi^2(t_1)\sin2\omega_p(t_1-t)+\\&-\frac{\lambda^2}{3}\int \frac{{\rm d}^3p{\rm d}^3q}{(2\pi)^6(8\omega_p \omega_q\omega_{p+q})}\int_0^t dt_1\Phi(t_1)\sin[\sum \omega_i)(t-t_1)]-\frac{\lambda^2}{2}\langle\phi^2\rangle^2\,\Phi(t)...=0,
\label{eq:2loopexplicit}
\end{flalign}
with $\sum \omega=\omega_p+\omega_q+\omega_{p+q}$. Here, the one-point function is evaluated over the non-squeezed coherent state \eqref{intro:coherent}.
Since the deviation of the field normalization $Z$ from unity is order $\hbar^2$, we introduce it only in front of 'classical' terms. All divergent quantum corrections of the first line correspond to the one-loop mass and coupling renormalization. Divergences of the second line correspond to the two-loop divergences.

In particular, the divergences of the first term of the second line are the analog of the sunset divergences, obtained in the standard vacuum analysis. 
To extract the divergent contributions of this integral, we integrate it by part three times.
With the first integration by part, we obtain
\begin{flalign}
   &\frac{\lambda^2}{3}\int \frac{{\rm d}^3p{\rm d}^3q}{(2\pi)^6(8\omega_p\omega_q\omega_{p+q})}\int_0^tdt_1\Phi(t_1)\sin[\sum \omega_i(t-t_1)]=\nonumber\\
  &\frac{\lambda^2}{3}\int \frac{{\rm d}^3p{\rm d}^3q}{(2\pi)^6(8\omega_p\omega_q\omega_{p+q})}\frac{1}{\sum \omega_i}  \left\{\Phi(t)-\Phi(0)\cos\left(\sum \omega_i t\right)- \int_0^tdt_1\dot{\Phi}(t_1)\cos(\sum \omega_i )(t-t_1)] \right\}.
\end{flalign}
The first term of the second line is the mass divergence generated by the sunset diagram. We can see this by evaluating the integral in dimensional regularization and checking that its divergences match the analogous ones one would get from the analysis of the effective action.
The second boundary term is the initial time singularity associated to the mass divergence. This term is finite as long as $t>0$ and it is divergent at the initial time, as we can see by directly evaluating it. 

The leftover time integral is still divergent and we integrate by part a second time
\begin{flalign}
   &-\frac{\lambda^2}{3}\int \frac{{\rm d}^3p{\rm d}^3q}{(2\pi)^6(8\omega_p\omega_q\omega_{p+q})}\int_0^tdt_1\dot{\Phi}(t_1)\cos[\sum \omega_i(t-t_1)]=\nonumber\\
  &\frac{\lambda^2}{3}\int \frac{{\rm d}^3p{\rm d}^3q}{(2\pi)^6(8\omega_p\omega_q\omega_{p+q})}\frac{1}{(\sum \omega_i)^2}  \left\{-\dot{\Phi}(0)\sin\left(\sum \omega_i t\right)- \int_0^t dt_1\ddot{\Phi}(t_1)\sin(\sum \omega_i )(t-t_1)] \right\}.
  \label{eq:lastPart}
\end{flalign}
The term which is proportional to $\dot{\Phi}$ converges for $t>0$ and it is null at initial time. Again, the residual time integral is still divergent and we perform the last integration by part
\begin{flalign}
  &-\frac{\lambda^2}{3}\int \frac{{\rm d}^3p{\rm d}^3q}{(2\pi)^6(8\omega_p\omega_q\omega_{p+q})}\frac{1}{(\sum \omega_i)^2}  \int_0^tdt_1 \ddot{\Phi}(t_1)\sin(\sum \omega_i )(t-t_1)]=\nonumber\\&-\frac{\lambda^2}{3}\int \frac{{\rm d}^3p{\rm d}^3q}{(2\pi)^6(8\omega_p\omega_q\omega_{p+q})}\frac{1}{(\sum \omega_i)^3}  \left\{\ddot{\Phi}(t)-\ddot{\Phi}(0)\cos\left(\sum \omega_i t\right)- \int_0^tdt_1 \dddot{\Phi}(t_1)\cos(\sum \omega_i )(t-t_1)] \right\}.
\end{flalign}
Finally, the leftover time integral is convergent. This occurs because performing an additional integration by parts would introduce boundary terms with sufficient powers of momenta in the denominator to ensure convergence.
In this case, only the first term is divergent, while the second term is convergent for $t>0$. 

If we regulate the divergent integral in dimensional regularization, we find
\begin{flalign}
  -\frac{\lambda^2 \ddot{\Phi}(t)}{3}\int \frac{{\rm d}^3p{\rm d}^3q}{(2\pi)^6(8\omega_p\omega_q\omega_{p+q})}\frac{1}{(\sum \omega_i)^3}=-\frac{\lambda^2}{12\epsilon(4\pi)^4}\ddot{\Phi}(t)+...
\end{flalign}
We see now that this divergence is the analog of the standard wavefunction divergence that occurs in the vacuum analysis. If we impose the prescription \eqref{eq:2loop_Z}, this specific divergence disappears.
Finally, the second boundary term of \eqref{eq:lastPart} is divergent if $t=0$. This corresponds to the initial time singularity resulting from the field renormalization.



\begin{thebibliography}{99}

\bibitem{Glauber:1963tx}
R.~J.~Glauber,
``Coherent and incoherent states of the radiation field,''
Phys. Rev. \textbf{131}, 2766-2788 (1963)
doi:10.1103/PhysRev.131.2766.


\bibitem{Sudarshan:1963ts}
E.~C.~G.~Sudarshan,
``Equivalence of semiclassical and quantum mechanical descriptions of statistical light beams,''
Phys. Rev. Lett. \textbf{10}, 277-279 (1963)
doi:10.1103/PhysRevLett.10.277.


\bibitem{Kibble:1965zza}
T.~W.~B.~Kibble,
``Frequency Shift in High-Intensity Compton Scattering,''
Phys. Rev. \textbf{138}, B740-B753 (1965)
doi:10.1103/PhysRev.138.B740.


\bibitem{Zhang:1990fy}
W.~M.~Zhang, D.~Feng and R.~Gilmore,
``Coherent States: Theory and Some Applications,''
Rev. Mod. Phys. \textbf{62}, 867-927 (1990)
doi:10.1103/RevModPhys.62.867.


\bibitem{Zhang:1999is}
W.~M.~Zhang,
``Coherent states in field theory,''
[arXiv:hep-th/9908117 [hep-th]].



\bibitem{Dvali:2011aa}
G.~Dvali and C.~Gomez,
``Black Hole's Quantum N-Portrait,''
Fortsch. Phys. \textbf{61}, 742-767 (2013)
doi:10.1002/prop.201300001
[arXiv:1112.3359 [hep-th]].


\bibitem{Dvali:2012en}
G.~Dvali and C.~Gomez,
``Black Holes as Critical Point of Quantum Phase Transition,''
Eur. Phys. J. C \textbf{74}, 2752 (2014)
doi:10.1140/epjc/s10052-014-2752-3
[arXiv:1207.4059 [hep-th]].


\bibitem{Dvali:2012wq}
G.~Dvali and C.~Gomez,
``Black Hole Macro-Quantumness,''
[arXiv:1212.0765 [hep-th]].


 
\bibitem{Dvali:2013vxa}
G.~Dvali, D.~Flassig, C.~Gomez, A.~Pritzel and N.~Wintergerst,
``Scrambling in the Black Hole Portrait,''
Phys. Rev. D \textbf{88}, no.12, 124041 (2013)
doi:10.1103/PhysRevD.88.124041
[arXiv:1307.3458 [hep-th]].

 

\bibitem{Dvali:2013eja}
G.~Dvali and C.~Gomez,
``Quantum Compositeness of Gravity: Black Holes, AdS and Inflation,''
JCAP \textbf{01}, 023 (2014)
doi:10.1088/1475-7516/2014/01/023
[arXiv:1312.4795 [hep-th]].


\bibitem{Dvali:2013lva}
G.~Dvali and C.~Gomez,
``Black Hole's Information Group,''
[arXiv:1307.7630 [hep-th]].


\bibitem{Berezhiani:2016grw}
L.~Berezhiani,
``On Corpuscular Theory of Inflation,''
Eur. Phys. J. C \textbf{77}, no.2, 106 (2017)
doi:10.1140/epjc/s10052-017-4672-5
[arXiv:1610.08433 [hep-th]].

 
\bibitem{Dvali:2017eba}
G.~Dvali, C.~Gomez and S.~Zell,
``Quantum Break-Time of de Sitter,''
JCAP \textbf{06}, 028 (2017)
doi:10.1088/1475-7516/2017/06/028
[arXiv:1701.08776 [hep-th]].


\bibitem{Dvali:2017ruz}
G.~Dvali and S.~Zell,
``Classicality and Quantum Break-Time for Cosmic Axions,''
JCAP \textbf{07}, 064 (2018)
doi:10.1088/1475-7516/2018/07/064
[arXiv:1710.00835 [hep-ph]].


\bibitem{Vachaspati:2017jtw}
T.~Vachaspati,
``Quantum Backreaction on Classical Dynamics,''
Phys. Rev. D \textbf{95}, no.12, 125002 (2017)
doi:10.1103/PhysRevD.95.125002
[arXiv:1704.06235 [hep-th]].


\bibitem{Kovtun:2020udn}
A.~Kovtun and M.~Zantedeschi,
``Breaking BEC: Quantum evolution of unstable condensates,''
Phys. Rev. D \textbf{105}, no.8, 085019 (2022)
doi:10.1103/PhysRevD.105.085019
[arXiv:2008.02187 [hep-th]].


\bibitem{Kovtun:2020ndc}
A.~Kovtun and M.~Zantedeschi,
``Breaking BEC,''
JHEP \textbf{07}, 212 (2020)
doi:10.1007/JHEP07(2020)212
[arXiv:2003.10283 [hep-th]].


\bibitem{Berezhiani:2020pbv}
L.~Berezhiani and M.~Zantedeschi,
``Evolution of coherent states as quantum counterpart of classical dynamics,''
Phys. Rev. D \textbf{104}, no.8, 085007 (2021)
doi:10.1103/PhysRevD.104.085007
[arXiv:2011.11229 [hep-th]].

\bibitem{Berezhiani:2021gph}
L.~Berezhiani, G.~Cintia and M.~Zantedeschi,
``Background-field method and initial-time singularity for coherent states,''
Phys. Rev. D \textbf{105}, no.4, 045003 (2022)
doi:10.1103/PhysRevD.105.045003
[arXiv:2108.13235 [hep-th]].

\bibitem{Dvali:2022vzz}
G.~Dvali and L.~Eisemann,
``Perturbative understanding of nonperturbative processes and quantumization versus classicalization,''
Phys. Rev. D \textbf{106} (2022) no.12, 125019
doi:10.1103/PhysRevD.106.125019
[arXiv:2211.02618 [hep-th]].


\bibitem{Michel:2023ydf}
M.~Michel and S.~Zell,
``The Timescales of Quantum Breaking,''
[arXiv:2306.09410 [quant-ph]].


\bibitem{Dvali:2015wca}
G.~Dvali and M.~Panchenko,
``Black Hole Type Quantum Computing in Critical Bose-Einstein Systems,''
[arXiv:1507.08952 [hep-th]].


\bibitem{Dvali:2018xpy}
G.~Dvali,
``A Microscopic Model of Holography: Survival by the Burden of Memory,''
[arXiv:1810.02336 [hep-th]].


\bibitem{Dvali:2020wft}
G.~Dvali, L.~Eisemann, M.~Michel and S.~Zell,
``Black hole metamorphosis and stabilization by memory burden,''
Phys. Rev. D \textbf{102}, no.10, 103523 (2020)
doi:10.1103/PhysRevD.102.103523
[arXiv:2006.00011 [hep-th]].

\bibitem{Eberhardt:2023axk}
A.~Eberhardt, A.~Zamora, M.~Kopp and T.~Abel,
``The classical field approximation of ultra light dark matter: quantum breaktimes, corrections, and decoherence,''
[arXiv:2310.07119 [astro-ph.CO]].


\bibitem{Berezhiani:2015ola}
L.~Berezhiani and M.~Trodden,
``How Likely are Constituent Quanta to Initiate Inflation?,''
Phys. Lett. B \textbf{749}, 425-430 (2015)
doi:10.1016/j.physletb.2015.08.007
[arXiv:1504.01730 [hep-th]].


\bibitem{Berezhiani:2022gnv}
L.~Berezhiani and M.~Trodden,
``A relativistic gas of inflatons as an initial state for inflation,''
Phys. Lett. B \textbf{840}, 137852 (2023)
doi:10.1016/j.physletb.2023.137852
[arXiv:2211.06222 [hep-th]].

\bibitem{Hertzberg:2016tal}
M.~P.~Hertzberg,
``Quantum and Classical Behavior in Interacting Bosonic Systems,''
JCAP \textbf{11}, 037 (2016)
doi:10.1088/1475-7516/2016/11/037
[arXiv:1609.01342 [hep-ph]].

\bibitem{Berges:2004yj}
J.~Berges,
``Introduction to nonequilibrium quantum field theory,''
AIP Conf. Proc. \textbf{739}, no.1, 3-62 (2004)
doi:10.1063/1.1843591
[arXiv:hep-ph/0409233 [hep-ph]].


\bibitem{Baacke:1996se}
J.~Baacke, K.~Heitmann and C.~Patzold,
``Nonequilibrium dynamics: A Renormalized computation scheme,''
Phys. Rev. D \textbf{55}, 2320-2330 (1997)
doi:10.1103/PhysRevD.55.2320
[arXiv:hep-th/9608006 [hep-th]].


\bibitem{Ai:2021gtg}
W.~Y.~Ai, M.~Drewes, D.~Glavan and J.~Hajer,
``Oscillating scalar dissipating in a medium,''
JHEP \textbf{11}, 160 (2021)
doi:10.1007/JHEP11(2021)160
[arXiv:2108.00254 [hep-ph]].


\bibitem{Pereira:2022lbl}
J.~Pereira and T.~Vachaspati,
``Stability analysis of non-Abelian electric fields,''
Phys. Rev. D \textbf{106}, no.9, 096019 (2022)
doi:10.1103/PhysRevD.106.096019
[arXiv:2207.05102 [hep-th]].


\bibitem{Vachaspati:2022gco}
T.~Vachaspati,
``Electric strings in non-Abelian theories,''
Phys. Rev. D \textbf{107}, no.3, L031903 (2023)
doi:10.1103/PhysRevD.107.L031903
[arXiv:2212.00808 [hep-th]].


\bibitem{Baacke:1997zz}
J.~Baacke, K.~Heitmann and C.~Patzold,
``On the choice of initial states in nonequilibrium dynamics,''
Phys. Rev. D \textbf{57}, 6398-6405 (1998)
doi:10.1103/PhysRevD.57.6398
[arXiv:hep-th/9711144 [hep-th]].


\bibitem{Baacke:1999ia}
J.~Baacke, D.~Boyanovsky and H.~J.~de Vega,
``Initial time singularities in nonequilibrium evolution of condensates and their resolution in the linearized approximation,''
Phys. Rev. D \textbf{63}, 045023 (2001)
doi:10.1103/PhysRevD.63.045023
[arXiv:hep-ph/9907337 [hep-ph]].


\bibitem{mukhanov}
V.~Mukhanov, ``Physical Foundations of Cosmology,'' Cambridge University Press, 2005, doi:10.1017/CBO9780511790553

\bibitem{Mukhanov:2007zz}
V.~Mukhanov and S.~Winitzki,
``Introduction to quantum effects in gravity,''
 Cambridge University Press, 2007,
 ISBN 978-0-521-86834-1, 978-1-139-78594-5.

\bibitem{Berezhiani:2021zst}
L.~Berezhiani, G.~Dvali and O.~Sakhelashvili,
``de Sitter space as a BRST invariant coherent state of gravitons,''
Phys. Rev. D \textbf{105}, no.2, 025022 (2022)
doi:10.1103/PhysRevD.105.025022
[arXiv:2111.12022 [hep-th]].


\bibitem{Kugo:1977zq}
T.~Kugo and I.~Ojima,
``Manifestly Covariant Canonical Formulation of Yang-Mills Field Theories: Physical State Subsidiary Conditions and Physical S Matrix Unitarity,''
Phys. Lett. B \textbf{73}, 459-462 (1978)
doi:10.1016/0370-2693(78)90765-7.


\bibitem{Kugo:1977yx}
T.~Kugo and I.~Ojima,
``Manifestly Covariant Canonical Formulation of Yang-Mills Field Theories. 1. The Case of Yang-Mills Fields of Higgs-Kibble Type in Landau Gauge,''
Prog. Theor. Phys. \textbf{60}, 1869 (1978)
doi:10.1143/PTP.60.1869.

\bibitem{Kugo:1977mk}
T.~Kugo and I.~Ojima,
``Manifestly Covariant Canonical Formulation of Yang-Mills Field Theories. 2. The Case of Pure Yang-Mills Theories Without Spontaneous Symmetry Breaking in General Covariant Gauges,''
Prog. Theor. Phys. \textbf{61}, 294 (1979)
doi:10.1143/PTP.61.294.


\bibitem{Kugo:1977mm}
T.~Kugo and I.~Ojima,
``Manifestly Covariant Canonical Formulation of Yang-Mills Field Theories. 3. The Case of Yang-Mills Fields of Higgs-Kibble Type in General Covariant Gauges,''
Prog. Theor. Phys. \textbf{61}, 644-655 (1979)
doi:10.1143/PTP.61.644.


\bibitem{Weinberg:1996kr}
S.~Weinberg,
``The quantum theory of fields. Vol. 2: Modern applications.''


\bibitem{LGO}
L.~Berezhiani, G.~Dvali, O.~Sakhelashvili, to appear.

\bibitem{Brahma:2020tak}
S.~Brahma, K.~Dasgupta and R.~Tatar,
``de Sitter Space as a Glauber-Sudarshan State,''
JHEP \textbf{02}, 104 (2021)
doi:10.1007/JHEP02(2021)104
[arXiv:2007.11611 [hep-th]].


 \end{thebibliography}
\end{document}